\documentclass[aps,prl,twocolumn,floatfix,superscriptaddress]{revtex4-2}
\pdfoutput=1
\usepackage{graphicx}
\usepackage{amsmath,amsthm}
\usepackage{amssymb}
\usepackage{braket}
\usepackage{xcolor}
\usepackage{bm}
\usepackage{caption}
\usepackage{subcaption}
\usepackage[bookmarks=false,linkcolor=blue,urlcolor=blue,colorlinks,citecolor=blue]{hyperref}
\usepackage[english]{babel}
\usepackage{esint}
\usepackage{braket}

\usepackage{graphicx}
\usepackage{epstopdf} 
\usepackage{pdfsync}
\usepackage[dvipsnames]{xcolor}
\usepackage[normalem]{ulem}
\usepackage{hyperref}
\usepackage{slashed}
\usepackage{dsfont}
\usepackage{float}%fix the position of figures%
\newcommand{\be}[1]{ \begin{eqnarray} \mbox{$\label{#1}$} }
   
\newcommand{\ee}{\end{eqnarray}}
\newcommand{\eeq}{\end{equation}}

\newcommand\hansc[1] {{\color{red}\emph{{#1}}}}

\begin{document}
% \pacs{xxx,xxx,xxx  }
% \title{Dynamical axial potential induced transverse current in Weyl Exciton Insulators}
\title{Sound Induced Hall Currents in Weyl Exciton Insulators}

\author{Xinhong Zhou}
\affiliation{Tsung-Dao Lee Institute, Shanghai Jiao Tong University, Shanghai, 201210, China}
\author{Varsha Subramanyan}
\email{varshas@sjtu.edu.cn}
\affiliation{Tsung-Dao Lee Institute, Shanghai Jiao Tong University, Shanghai, 201210, China}

\author{Qing-Dong Jiang}
\email{qingdong.jiang@sjtu.edu.cn}
\affiliation{Tsung-Dao Lee Institute, Shanghai Jiao Tong University, Shanghai, 201210, China}
\affiliation{School of Physics and Astronomy, Shanghai Jiao Tong University, Shanghai, 201210, China}

\begin{abstract}
Weyl semimetals exhibit anomalous transport controlled by their gapless chiral nodes, while gapped Weyl systems are often expected to lose such distinctive responses. Here we show that Weyl excitonic insulators instead host a new form of axial-field-driven transport that exists only in the massive phase. Starting from the low-energy Breit-type Hamiltonian for a gapped Weyl system coupled to strain-induced axial potentials, we develop a semiclassical wave-packet theory and identify a dissipative transverse current generated by a dynamical axial potential. This response is qualitatively distinct from both ordinary vector-potential transport in gapped systems and axial responses in gapless Weyl semimetals. Physically, it originates from a mass-induced Berry structure of the reconstructed Weyl bands, which becomes active when the system is driven out of equilibrium by a chiral chemical-potential imbalance. We show that transverse sound waves provide a natural route to generate the required dynamical axial field and estimate the resulting current for realistic material parameters. Our results reveal sound-induced Hall transport as a direct probe of Weyl excitonic order and provide a transport signature of interaction-generated mass in Weyl materials. 

\end{abstract}
                                        
\maketitle   

\emph{Introduction.\textemdash}Weyl semimetals are 3D topological materials characterized by gapless Weyl points in their band spectra \cite{Armitage2018, Yan2017,Lv2015,Xu2015}. The chiral and topological nature of their low-energy  response makes them suitable platforms to study the physical effects of quantum anomalies \cite{Zyuzin2012,Ong2021}. Further, their response to strain can be modeled as a coupling to axial potentials, thus providing a system with mixed anomalies \cite{Cortijo2015,Pikulin2016}.
A variety of anomaly-induced transport phenomena have been proposed and studied in this context~\cite{Huang2017}, with much of the existing literature focusing on the topological ``$\theta$-term'' in the effective action. In this work, we show that a mass term, interpreted as an excitonic gap, also leads to novel, and distinct, transport phenomena.

\begin{figure}[t]
  \centering
  \begin{minipage}[t]{\linewidth}
    \raggedright  % 左对齐
    \textbf{(a)}\\  % 第一行：标签在左上角
    \vspace{2pt}    % 标签与图片之间的间距
    \centering      %图片居中
    \includegraphics[width=\linewidth]{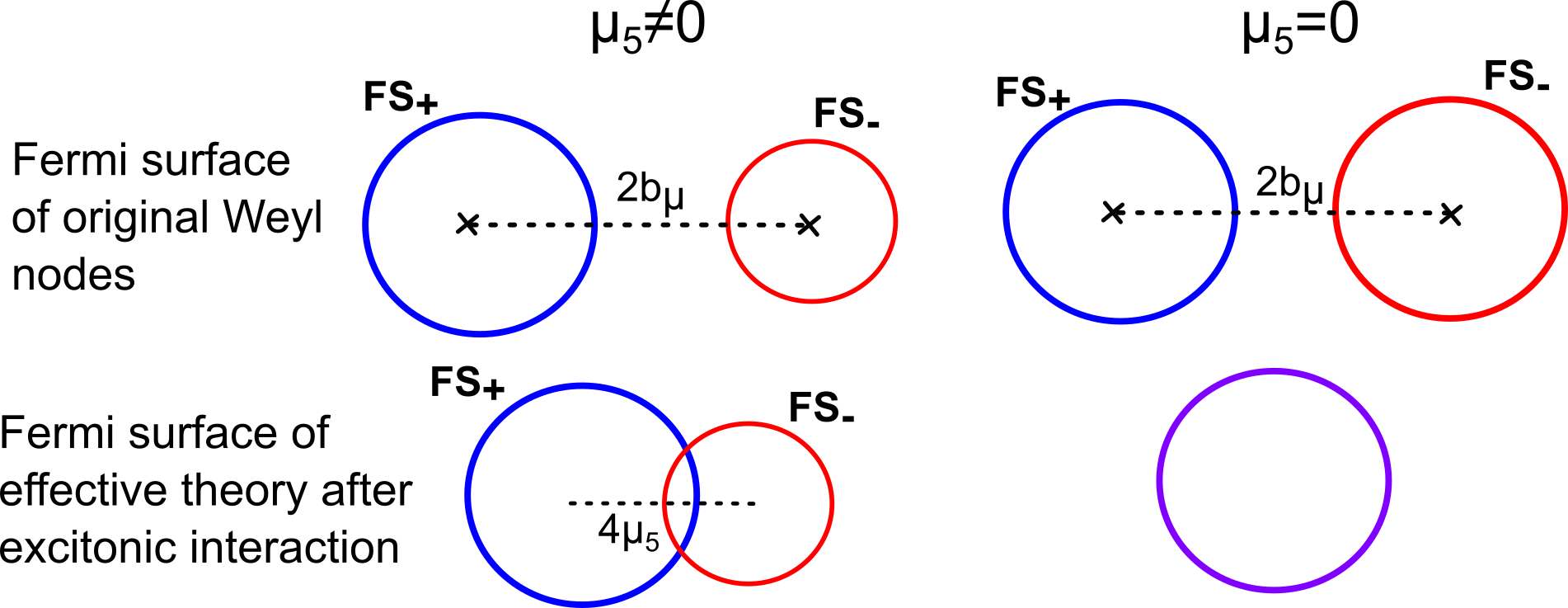}
    \label{WEI_cartoon}
  \end{minipage}
  \hfill
  \begin{minipage}[t]{0.48\linewidth}
    \raggedright  % 左对齐
    \textbf{(b)}\\  % 第一行：标签在左上角
    \vspace{2pt}    % 标签与图片之间的间距
    \centering      %图片居中
    \includegraphics[width=\linewidth]{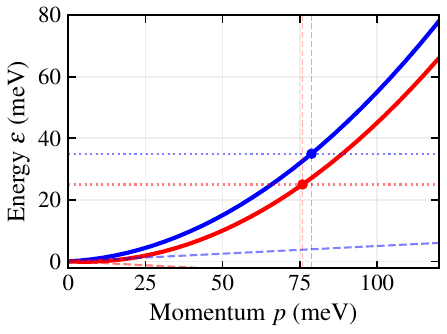}
    \label{band_structure_m_100meV,mu5=5meV,2columns.pdf}
  \end{minipage}
  \hfill
  \begin{minipage}[t]{0.48\linewidth}
    \raggedright
    \textbf{(c)}\\  % 第一行：标签在左上角
    \vspace{2pt}
    \centering
    \includegraphics[width=\linewidth]{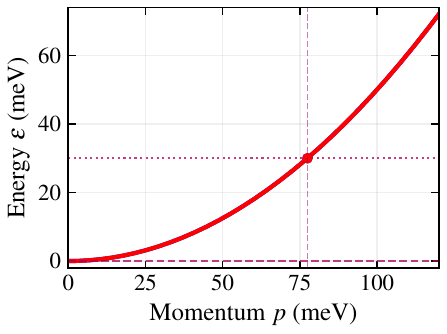}
    \label{band_structure_m_100meV,mu5=0meV,2columns.pdf}
  \end{minipage}
  \hfill
  
  \caption{(a) Illustration of the excitonic interaction in Weyl materials giving rise to the two-band model in $H_0(\mathbf{p})$. The band structure for $H_0(\mathbf{p})$ at $\mu=30\;\mathrm{meV}$ and $m=100\;\mathrm{meV}$ is shown here for (b) $\mu_5=5\;\mathrm{meV}$ and (c) $\mu_5=0\;\mathrm{meV}$. Red (blue) color represents $s=+(-)$. Solid curves represent $s$ band; dashed lines represent tangent line for $s$ band at $p=0$; dotted horizontal lines represent the $\mu_{s}$ intersecting with $s$ band at $p^*_s$. }
  \label{fig: band_structure_m_100meV}
\end{figure}

Strong excitonic interactions can couple Weyl nodes of opposite chirality and open a gap, producing a Weyl excitonic insulator (WEI). Such phases have attracted recent interest as platforms for studying interaction-driven topological phase transitions and unconventional transport responses~\cite{Grigoreva2025,Chang2021,Singh2018,Tabert2016,Behrends2019}. In the accompanying paper, Ref.~\onlinecite{longpaper}, we derived the effective action and the corresponding low-energy Breit-type Hamiltonian for the WEI phase in the large-mass limit. Below we summarize the results for completeness and then use the modified band structure to calculate transport. 
Figure~\ref{fig: band_structure_m_100meV}(a) schematically illustrates the reconstructed Fermi surface, while Figs.~\ref{fig: band_structure_m_100meV}(b) and~\ref{fig: band_structure_m_100meV}(c) show the effective band structures for different values of the axial potential, using realistic material parameters adopted in this study.

Using a semiclassical wave-packet and Boltzmann approach, generalized from Refs.~\onlinecite{Son2013,Niu1999} to massive Weyl bands coupled to both vector and axial potentials, we show that a dynamical axial potential can induce a transverse response current in a WEI. This current appears when the gapped Weyl bands are driven out of equilibrium and acquire a chemical-potential imbalance. The effect is qualitatively different from the response to ordinary vector potentials in gapped systems, and also from the axial response of gapless Weyl semimetals. We analyze the dependence of the transverse current on the excitonic mass and on the driving frequency, and show how the response changes when the chemical-potential imbalance is absent. Finally, we estimate the magnitude of the effect using realistic material parameters. The resulting transverse current provides a potential transport signature of the Weyl excitonic phase and offers a possible way to probe the size of the excitonic gap.

\emph{Model construction.\textemdash} In Ref. \onlinecite{longpaper}, starting from the Dirac equation, we derived a two-component low-energy field theory for massive Dirac fermions coupled to both vector and axial potentials.
Here, we summarize the ingredients needed for the transport calculation in the Weyl excitonic insulator (WEI) phase.

We begin with an excitonic interaction between left- and right-handed Weyl nodes,
\be{exciton pairing between two Weyl nodes}
\hat H_{int}=-\frac{1}{V}\sum_{\mathbf{k,k',q}}U(\mathbf{q})( c^\dagger_{\mathbf{k+q}L} c_{\mathbf{k}R})(c^\dagger_{\mathbf{k'}R} c_{\mathbf{k'+q}L})
\ee
which opens a gap $\Delta(x)\equiv m\exp[i\theta(x)]$. This gap can be represented as a complex mass term in the Dirac Lagrangian~\cite{Zyuzin2012}:
\be{Lagrangian for Weyl semimetal with coupling effect between left and right Weyl nodes}
{\cal L}=\bar\psi\left\{i\gamma^\mu[\partial_\mu+iA_\mu+i\gamma^5(\tilde A^5_\mu+b_\mu)]-m e^{i\gamma^5\theta(x)}\right\}\psi.\nonumber\\
\ee
Here $A_\mu$ and $\tilde A^5_\mu$ are the external vector and axial potentials, respectively, while $b_\mu$ denotes the separation of the two Weyl nodes in energy-momentum space. Performing the chiral rotation $\psi\rightarrow e^{-i\gamma^5\theta(x)/2}\psi$, and keeping the anomalous Jacobian \iffalse required to preserve the vector current $j^\mu$\fi, gives the effective action~\cite{Huang2017,Yu2021}
\be{equivalence action to describe the system}
\Gamma[A_\mu,A^5_\mu]&=&\int_x\bar\psi[i\gamma^\mu(\partial_\mu+iA_\mu+i\gamma^5 A^5_\mu)-m]\psi\nonumber\\
&-&\int_x\frac{ \theta}{64\pi^2}\epsilon^{\mu\nu\alpha\beta}(F_{\mu\nu}^{+}F_{\alpha\beta}^{+}+F_{\mu\nu}^{-}F_{\alpha\beta}^{-})
\ee
where $A^5_\mu\equiv\tilde A^5_\mu+b_\mu-\partial_\mu\theta/2$ is the effective axial potential and $F_{\mu\nu}^{\pm}\equiv\partial_\mu(A_\nu\pm A^5_\nu)-(\mu\leftrightarrow\nu)$. 
In what follows, we focus on the first term in Eq.\eqref{equivalence action to describe the system}, which captures the mass-dependent coupling between vector and axial fields and is responsible for the transport response studied here.

Writing the Dirac spinor in terms of upper and lower components, $(\phi,\chi)^{\mathrm{T}}$, the massive Dirac Lagrangian becomes
\be{Dirac Lagrangian in two component version}
{\cal L}=\phi^\dagger(i\bar D_0)\phi &+&\chi^\dagger(i\bar D_0+2m)\chi\nonumber\\ 
&+&\phi^\dagger(i\sigma^i\bar D_i)\chi+\chi^\dagger(i\sigma^i\bar D_i)\phi
\ee
where $\bar D_0=\partial_0+ieA_0+i\eta\sigma^i A^5_i$ and $\bar D_i=\partial_i+ieA_i-\frac{i}{3}\eta\sigma_i A^5_0$. Here $e$ and $\eta$ denote the vector and axial charges, respectively. These modified covariant derivatives show that axial fields couple to a massive Weyl system in a way that is qualitatively different from the gapless case. After integrating out the negative-energy component $\chi$, the low-energy effective Lagrangian to $O(m^{-2})$ is
\be{effective Lagrangian 1}
{\cal L}_{eff}&=&\phi^\dagger\bigg\{(i\bar D_0)+\frac{1}{2m}(\sigma^i\bar D_i)^2\nonumber\\
&+&\frac{-1}{4m^2}(\sigma^i\bar D_i)(i\bar D_0)(\sigma^j\bar D_j)\bigg\}\phi
\ee
We now rescale the field $\phi$ to $\phi=\left[1+\frac{1}{8m^2}(\sigma^i\bar D_i)^2\right]\tilde\phi$ to ensure standard commutation relations,  and recast the effective Lagrangian so that the equations of motion is in  Schr\"odigner form: $(i\partial_t-\hat{H}_{eff})\tilde{\phi}=0$. Expanding in orders of $1/m$, we then obtain
\be{effective Hamiltonian after rescaling}
\hat H^{(0)}_{eff} &=& eA_0+\eta A^5_i\sigma^i\nonumber\\
\hat H^{(1)}_{eff} &=& \frac{1}{2m}(\bm{\hat\pi}^2+\mu_5^2)+\frac{1}{2m}(-eB_i+2\mu_5\hat\pi_i)\sigma^i\nonumber\\
\hat H^{(2)}_{eff} &=& \frac{e}{8m^2}[-\bm{\nabla}\cdot\mathbf{E}+(\bm{\hat\pi}\times\mathbf{E}-\mathbf{E}\times\bm{\hat\pi})\cdot\bm{\sigma}]\nonumber\\
&+&\left(\frac{-\eta}{4m^2}\right)(\mathbf{B_5}\cdot\bm{\hat\pi})+\frac{e\eta}{2m^2}(B_iA^5_i)
\ee
where $\hat\pi_i\equiv\hat p_i+eA_i=-i\partial_i+eA_i$ and $\mu_5\equiv\eta A^5_0$ is the chiral chemical potential that we set constant in the following discussion. Also, we assume $|(\mathbf{b}-\bm{\nabla}\theta/2)|/|\tilde{\mathbf{A}}^5|\simeq0$ so we can view $\mathbf{A}^5\simeq\tilde{\mathbf{A}}^5$ for simplicity \cite{Zyuzin2012,Yu2021}. Eq.\eqref{effective Hamiltonian after rescaling} amounts to a generalization of the Breit Hamiltonian that includes axial fields. A similar Hamiltonian was obtained in \cite{SHAPIRO2002113} using the Foldy-Wouthuysen transformation in the context of space-time torsion.%, though  dynamical axial potentials are more easily generated in condensed matter systems.

{\it Wavepacket method and semi-classical transport.\textemdash} To calculate response in the presence of dissipation, we will use a semi-classical phase space analysis at zero temperature \cite{Son2013,Niu1999,JiangQD2016,Medel2023}. The analysis applies in the regime $K_BT\ll\mu_5$, where thermal fluctuations do not appreciably mix the $s=\pm$ bands or generate higher-order corrections~\cite{Gorbar2017}. While transport induced by vector and axial fields in gapless Weyl materials is well understood, our goal here is to show how the interplay of these fields leads to qualitatively distinct responses once the Weyl spectrum is gapped~\cite{supp}. 

Assuming that the applied fields $\{A_\mu,A^5_i\}$ are small ($O(\varepsilon)$) and vary slowly in space and time ($O(\delta)$), we start from the non-perturbative Hamiltonian from Eq.\eqref{effective Hamiltonian after rescaling}:
\be{non-perturbative Hamiltonian}
H_0(\bm{\pi})=\frac{\bm{\pi}^2}{2m}+\frac{\mu_5}{m}\bm{\pi}\cdot\bm{\sigma}
\ee

The bands corresponding to $H_0(\mathbf{p})$ are shown in Figure \ref{fig: band_structure_m_100meV}. In the effective theory, it's $\mu_5$, rather than the Weyl-node separation $\mathbf{b}$, that separates the two renormalized bands and their Fermi surfaces. For the realistic parameters used below, one finds
\be{valid the choice of H_0(p)}
\frac{\eta\hbar v_FA^5_i}{(\mu_5p_s^*/m)}\lesssim O(10^{-4}).
\ee
It is therefore justified to treat the term $\eta A^5_i\sigma^i$, together with the remaining field-dependent terms in Eq.~\eqref{effective Hamiltonian after rescaling}, as a perturbation relative to $\mu_5\mathbf{p}\cdot\bm{\sigma}/m$. The energy of a wavepacket centered at $({\bf r_c},{\bf p_c})$ is evaluated using the expanded Hamiltonian to be \cite{supp} 
\begin{align}
\begin{split}
    \epsilon_{s}(&\mathbf{r}_c,\mathbf{p}_c)=\left(\frac{|\mathbf{p}_c|^2}{2m}+\frac{s\mu_5|\mathbf{p}_c|}{m}\right)-\frac{e\mu_5B_jp_{cj}}{2m|\mathbf{p}_c|^2}\\ &+s\left(\frac{\eta A^5_jp_{cj}}{|\mathbf{p}_c|}-\frac{eB_jp_{cj}}{2m|\mathbf{p}_c|}\right)+\left(\frac{e\eta}{2m^2}B_jA^5_j-\frac{\eta}{2m^2}B_{5j}p_{cj}\right)
\end{split}
\end{align}
where $s=\pm 1$ labels the two bands. This modified spectrum amounts to a redistribution  of occupied states into the phase space. In the following calculation we scale the definition of $A^5_i$ by $\eta/e$ for simplicity of notation. Recalling that the applied fields are small ($O(\varepsilon)$) and vary slowly in space and time ($O(\delta)$), the modified distribution function $n_s$ can be  obtained from the Boltzmann transport equation
\begin{align}
    \begin{split}
        \partial_t n_s + {\bf \dot{r}}_c\cdot\partial_{\bf {r}_c}n_s +\dot{\mathbf{p}}_c\cdot\partial_{\mathbf{p}_c}n_s = I[n_s]
    \end{split}
\end{align}
where $I[n_s]$ is the collision integral, and ${\bf \dot{p}}_c$ and ${\bf \dot{r}}_c$ are the standard wavepacket equations of motion - 
\begin{align}
        \mathbf{\dot p}_c=-\frac{\partial\epsilon_s}{\partial\mathbf{r}_c}+e(\mathbf{E}+\mathbf{\dot r}_c\times\mathbf{B}),\quad
\mathbf{\dot r}_c=\frac{\partial\epsilon_s}{\partial\mathbf{p}_c}+\mathbf{\dot p}_c\times\bm{\Omega}_s
\end{align}
Here $\bm{\Omega}_s =  s\widehat{\mathbf{p}} /2|\mathbf{p}|^2$ is the Berry curvature of the $s$-band. To obtain corrections to the electron distribution function $n_s$ up to orders $O(\varepsilon)$ and $O(\varepsilon\delta)$, we solve the modified Boltzmann transport equation 
\begin{align}
\begin{split}
(\partial_t+\mathbf{v}_{cs}\cdot\partial_{\mathbf{r}_c})n_s&+\left(e\mathbf{E}-\frac{\partial\epsilon_s}{\partial\mathbf{r}_c}+e\mathbf{v}_{cs}\times\mathbf{B}\right)\frac{\partial n_s}{\partial\mathbf{p}_c}\\&=-\frac{1}{\tau_0}[\delta n_s^{(\varepsilon)}+\delta n_s^{(\varepsilon\delta)}]
\end{split} \, ,
\end{align}
where $\delta n_s^{(x)}$ denotes corrections to the electron distribution of order $O(x)$. We have introduced a phenomenological relaxation time $\tau_0$ to account for dissipation.

The total response current in the Boltzmann formalism is \cite{Son2013} 
\begin{align}
\begin{split}
\mathbf{j}_s(\mathbf{r}_c,t)=e\int_{\bf{p}_c}\bigg[-\epsilon_s\frac{\partial n_s}{\partial\mathbf{p}_c}&-e\left(\bm{\Omega}_s\cdot\frac{\partial n_s}{\partial\mathbf{p}_c}\right)\epsilon_s\mathbf{B}\\&-\epsilon_s\left(\bm{\Omega}_s\times\frac{\partial n_s}{\partial\mathbf{r}_c}\right)\bigg] \, .
\end{split}
\end{align}
The three terms correspond, respectively, to the conduction current, the chiral magnetic contribution, and the three-dimensional Hall contribution. Substituting the corrected distribution function into the current gives~\cite{supp}
\be{final expression of response current}
    j_{s,i}^{(\varepsilon)}(\mathbf{q},\omega)&=&(ie^2{p^*_s}^2v^*_s)I_{ij,s}(\mathbf{q},\omega)E_{s,j}^{(\varepsilon)}\label{varepsilon order response current}\\
j_{s,i}^{(\varepsilon\delta)}(\mathbf{q},\omega)&=&\frac{s\mu_s e^2B^{(0)}_i}{4\pi^2}-\frac{e^2{p^*_s}\mu_5}{2m}\omega I_{ij,s}(\mathbf{q},\omega)B^{(0)}_j\nonumber\\
&-&\frac{se^2{p^*_s}^2}{2m}\omega I_{ij,s}(\mathbf{q},\omega)B^{(\varepsilon\delta)}_{s,j}\nonumber\\
&+&\Theta(|\mu_5|)se^2\mu_s\epsilon_{ijk}q_kI_{jl,s}(\mathbf{q},\omega)E^{(\varepsilon)}_{s,l}\label{varepsilon-delta order response current} \, ,
\ee
where $\Theta(x)$ is the step function: $\Theta(x)=1$ for $x>0$ and otherwise vanishes. Here we define the effective wavepacket momentum $p_s^{*}=-s\mu_5+\sqrt{\mu_5^2+2m\mu_s}$, effective wavepacket velocity 
$v_s^{*}=\left(\sqrt{\mu_5^2+2m\mu_s}\right)/{m}$, and the Fermi surface integral: 
\be{angular integrals for the above equations}
I_{ij,s}(\mathbf{q},\omega)&=&\int\frac{d\widehat{\mathbf{p}}_c}{(2\pi)^3}\frac{\widehat{p}_{ci}\widehat{p}_{cj}}{\omega-v_s^{*}\cdot\mathbf{q}\cdot\widehat{\mathbf{p}}_c+i\tau_0^{-1}}
\ee
\iffalse\hansc{If I understand correctly, in the limit you are using you simply put, $$ I_{ij,s}(\mathbf{q},\omega) = \frac {-i\tau_0} {2\pi^2}$$ if this is correct I think you should write it explicitly.}\fi
Note that the effective electromagnetic fields experienced by the wavepacket  \cite{supp}
\begin{align}\label{modified EM fields}
\begin{split}
E_{s,i}^{(\varepsilon)}&=-\left[i\omega\left(\frac{s}{v_s^{*}}A^5_i+A_i\right)+iq_iA_0\right]=E_i^{(0)}-i\omega\frac{s}{v_s^{*}}A^5_i\\
B_{s,i}^{(\varepsilon\delta)}&=-i\epsilon_{ikl}q_k\left(\frac{sp^*_s}{m}A^5_l+A_l\right)=B^{(0)}_i-i\epsilon_{ikl}q_k\frac{sp^*_s}{m}A^5_l
\end{split}
\end{align}
are explictly mass-dependent. This is qualitatively different from the gapless Weyl-semimetal case, where axial fields enter through the simple chirality-dependent replacement $A_i\rightarrow A_i+\lambda A^5_i$, with $\lambda$ denoting the chirality of the Weyl node.
Further, we see that the the current $j^{(\epsilon\delta)}_{s,i}$ contains the topological transport currents.  The first term corresponds to the chiral magnetic effect  \cite{Fukushima2008}, the second  gives corrections to the chiral magnetic effect due to the mass $m$, and the third  contains corrections due to both mass and the axial fields. Similarly, the last term in the current shows the modified 3d Hall effect. It is notable that these effects are most clearly seen when the system is in a non-equilibrium state in the presence of the term $\mu_5\neq 0$. 

\begin{figure}[t]
  \centering
  % 第一行：(a) 和 (b)
  \begin{minipage}[t]{0.48\linewidth}
    \raggedright
    \textbf{(a)}\\
    \vspace{2pt}
    \centering
    \includegraphics[width=\linewidth]{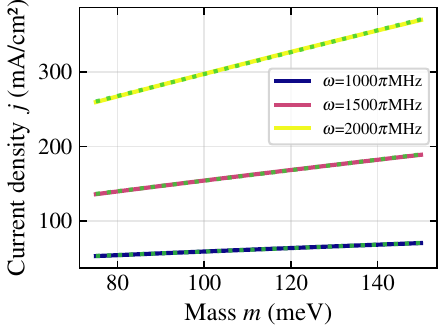}
    \label{j_vs_m_fixed_omega_linear_with_fit_mu5=5meV,2columns.pdf}
  \end{minipage}
  \hfill
  \begin{minipage}[t]{0.48\linewidth}
    \raggedright
    \textbf{(b)}\\
    \vspace{2pt}
    \centering
    \includegraphics[width=\linewidth]{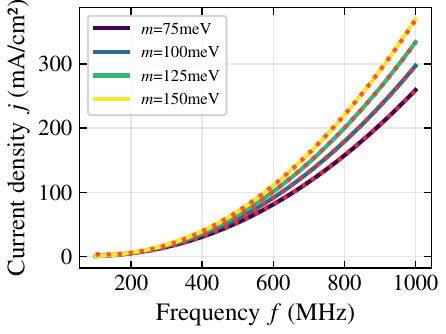}
    \label{j_vs_omega_fixed_m_linear_with_fit,mu5=5meV,2columns.pdf}
  \end{minipage}
  
  \vspace{6pt}  % 两行之间的垂直间距
  
  % 第二行：(c) 和 (d)
  \begin{minipage}[t]{0.48\linewidth}
    \raggedright
    \textbf{(c)}\\
    \vspace{2pt}
    \centering
    \includegraphics[width=\linewidth]{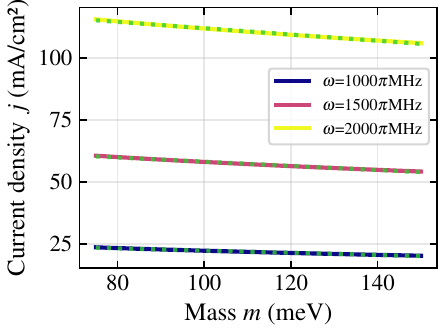}
    \label{j_vs_m_fixed_omega_linear_with_fit_mu5=0meV,2columns.pdf}
  \end{minipage}
  \hfill
  \begin{minipage}[t]{0.48\linewidth}
    \raggedright
    \textbf{(d)}\\
    \vspace{2pt}
    \centering
    \includegraphics[width=\linewidth]{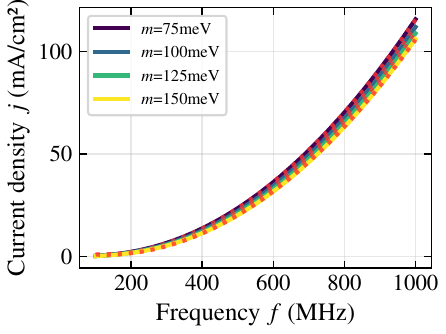}
    \label{j_vs_omega_fixed_m_linear_with_fit,mu5=0meV,2columns.pdf}
  \end{minipage}
  \begin{minipage}[t]{\linewidth}
    \raggedright
    \textbf{(e)}\\  % 第一行：标签在左上角
    \vspace{2pt}
    \centering
    \includegraphics[width=0.7\linewidth]{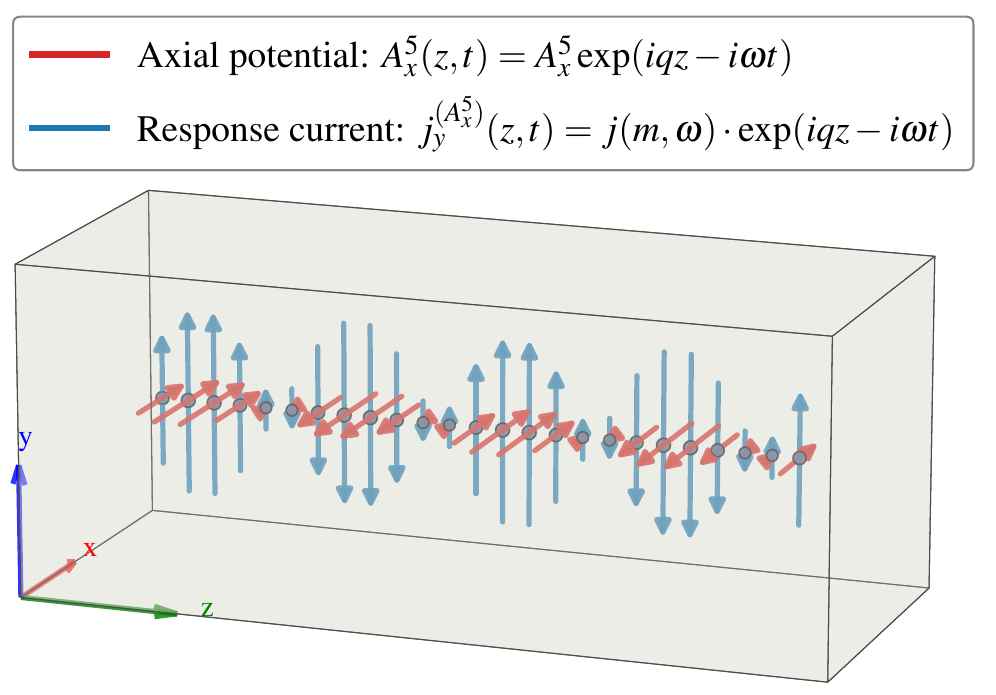}
    \label{fig: Hall-type response current induced by S_x(z,t)}
  \end{minipage}
  \caption{Analysis of transverse response current density at $\mu=30\;\mathrm{meV},m=100\;\mathrm{meV}$.  Relationship between $j(m,\omega)$ and $m$ with fixed $\omega$ at (a) $\mu_5=5\;\mathrm{meV}$ and (c) $\mu_5=0\;\mathrm{meV}$.  Relationship between $j(m,\omega)$ and $f=\omega/2\pi$ with fixed $m$ at (b) $\mu_5=5\;\mathrm{meV}$ and (d) $\mu_5=0\;\mathrm{meV}$. Dotted lines represent fit data. (e) Schematic figure showing the transverse response current $j_y^{(A^5_x)}$ induced by the dynamical axial potential $A^5_x(z,t)$.}
  \label{fig: Hall-type response current induced by S_x(z,t)}
\end{figure}

\emph{Transverse response current induced by dynamical axial potential.\textemdash} We now  highlight the contributions from the last two terms in Eq.\eqref{varepsilon-delta order response current} - 
\be{Transverse response current induced by dynamical axial potential}
    j_{s,i}^{(\varepsilon\delta)}(\mathbf{q},\omega)&=&\Theta(|\mu_5|) se^2\mu_s\epsilon_{ijk}q_kI_{jl,s}(\mathbf{q},\omega)E^{(\varepsilon)}_{s,l}\nonumber\\
&-&\frac{se^2{p^*_s}^2}{2m}\omega I_{ij,s}(\mathbf{q},\omega)B^{(\varepsilon\delta)}_{s,j}
\ee
We pick a dynamical monochromatic axial potential of the form $A^5_x(z,t)=A^5_x\exp(iqz-i\omega t)$ and set $A_\mu=0$. This ensures that the effective electromagnetic fields in Eq.\eqref{modified EM fields} are only due to the dynamical axial potential. Thus, it is the axial potential that drives transport in this setup. For simplicity we choose $\mathbf{q}=(0,0,q)$, chemical potential $\mu\simeq30\;\mathrm{meV}$ \cite{Weng2015}, chiral chemical potential $\mu_5\simeq5\mathrm{\;meV}$ and the strength of exciton pairing $m\in[75,150]\;\mathrm{meV}$ \cite{Zhang2024}. In Weyl semi-metal systems, the typical value of Fermi velocity is $v_F\simeq 10^5\;\mathrm{m/s}$ and the relaxation time $\tau_0\simeq 10^{-11}\;\mathrm{s}$. In experiments, $A^5_x(z,t)$ can be induced by a transverse sound wave with the linear dispersion relationship $\omega=v_0 q$ ($v_0\simeq 10^3\mathrm{\;m/s}$ is the sound velocity) where $\omega/2\pi\in[10^2,10^3]\;\mathrm{MHz}$ and $A^5_x(z,t)\simeq e^{-1}b_zu_0q\exp(iqz-i\omega t)$ \cite{Huang2017,Liang2025,Chernodub2019}. Here $b_z\simeq 0.2\pi a^{-1}$ is the separation of Weyl nodes in $k_z$ direction and $u_0\simeq 0.01a$ is the displacement amplitude ($a$ is the lattice constant). In this condition, we have $|\mathrm{Re}[I_{yx,s}(q,\omega)]/\mathrm{Im}[I_{yx,s}(q,\omega)]|<0.02$ so the total transverse response current summing up both $s=\pm$ induced by $A^5_x(z,t)$ reads: $j_y^{(A^5_x)}(z,t)\equiv j(m,\omega)\exp(iqz-i\omega t)$. Here $j(m,\omega)$ is the magnitude of response current. To understand the dependence of $j(m,\omega)$ on mass $m$ and frequency $\omega$, we scan the region $m\in[75,150]\;\mathrm{meV},\;f=\omega/2\pi\in[10^2,10^3]\;\mathrm{MHz}$.

The total transverse current under these parameters is shown in Fig.~\ref{fig: Hall-type response current induced by S_x(z,t)}. We see a clear positive sloped linear dependence  of the current density on the mass term $m$, and a quadratic dependence on frequency $\omega$ when $\mu_5\neq 0$ in Fig.~\ref{fig: Hall-type response current induced by S_x(z,t)}(a) and (b). On the other hand, when $\mu_5=0$, the slope of the mass dependence becomes negative while the frequency dependence remains quadratic as seen in Fig.~\ref{fig: Hall-type response current induced by S_x(z,t)}(c) and (d). Such a transverse response is absent in gapless systems entirely, and are also not induced by the vector potentials alone, making this current a unique feature of transport in the presence of axial potentials. 

The mass and frequency dependence can be understood by a simple order counting using Eqs.\eqref{angular integrals for the above equations}-\eqref{Transverse response current induced by dynamical axial potential}  and the expressions of $\{A^5_x(z,t),p_s^*,v_s^*\}$. The second term of $j_{s,i}^{(\varepsilon\delta)}$ is $O(m^0\omega^2)$ while the first term is $O(m\omega^2)$. When $\mu_5=0$ only the second term, originating from $(-\eta/4m^2)(\mathbf{B}_5\cdot\bm{\hat\pi})$ in the effective Hamiltonian $\hat H_{eff}^{(2)}$, contributes to the transverse current. In this condition, $j(m,\omega)$ decreases with increasing $m$, and has quadratic dependent on $\omega$. After turning on $\mu_5=5\;\mathrm{meV}$ the first term in Eq.\eqref{Transverse response current induced by dynamical axial potential} which comes from spin-orbit interaction between nonzero Berry connection $\bm{\Omega}_s$ and $(\eta A^5_i\sigma^i)$ in the effective Hamiltonian $\hat H_{eff}^{(0)}$ is the dominant contribution to $j_y^{(A^5_x)}$. So $j(m,\omega)$ has a positive slope with respect to $m$ now and remains quadratic in $\omega$.

Physically, the Hall-type transverse response obtained in this work is due to the existence of a non-trivial Berry connection in the modified band structure of the gapped system. This is seen in the $\mu_5\mathbf{p}\cdot\bm{\sigma}/m$ term in Hamiltonian $H_0({\bf p})$, which is present whenever $\mu_5\neq 0$.   This Berry connection is distinct from that of the original Weyl nodes, which have now been gapped out. Instead, it arises from a net spin-orbit coupling in the system due to an imbalance in the chemical potential in the original Weyl nodes, and thus the resulting effective Hamiltonian. It comes from the modification of the positive-energy band $\phi$ due to the integrating out of the negative-energy band $\chi$, and contains contributions from both the original left-movers and right-movers. The gapped system thus retains its topology when it is driven out of equilibrium. This idea is further emphasized when considering the band velocity: $(\partial\epsilon_s^{(0)}/\partial \mathbf{p})=\mathbf{p}/m+(s\mu_5/m)\widehat{\mathbf{p}}$. That is, we obtain a chiral band structure at the origin when $\mu_5\neq 0$. This structure clearly vanishes when $\mu_5=0$ and there is no chemical potential imbalance.  

\emph{Experimental realization \textemdash}
The transport response identified in this work requires three ingredients: a Weyl material capable of entering an excitonic insulating phase, a nonequilibrium Fermi-surface imbalance, and a slowly varying sound wave. We propose the following setup. First, a voltage or current bias is applied along the symmetry axis of the material, which we take to be the $z$ direction, together with a parallel magnetic field. Such an electromagnetic configuration generates a chiral chemical-potential imbalance, $\mu_5\propto {\bf E\cdot B}$~\cite{Spivak2013}. Once the material is tuned into the WEI phase, for example by doping, chemical-potential tuning, or static strain~\cite{Singh2018,Grigoreva2025,Wei2012}, a transverse sound wave is applied in the plane perpendicular to the symmetry axis. As discussed above, this sound wave generates a dynamical axial potential and drives a Hall-type current. The resulting transverse voltage across the sample provides an experimentally accessible signature of the response.

Since our results are  direct consequences of the interplay between axial and vector potentials in massive Breit Hamiltonians, it is appropriate to compare the relative magnitudes of this response to that of EM only in terms of their respective structure constants $\alpha_5=\eta^2/4\pi$ and $\alpha=e^2/4\pi$. In real materials where we expect strain to couple to the band structure as an axial potential (like in Dirac and Weyl semimetals), $\eta$ is related to the Grun\"{e}issen parameter $\beta$ \cite{Cortijo2015}. This dimensionless parameter is obtained by thermodynamic considerations and is a measure of the deformability of the lattice potential in response to external strain. While  this  quantity is material dependent, it is reasonable to use estimates for 3D materials with Lennard-Jones potentials where $\beta\sim 19/6$. Since $\alpha_5\sim O(\beta^2)$, the axial-field-induced current is expected to be at least comparable to, and possibly larger than, the corresponding electromagnetic response. The currents shown in Fig.~\ref{fig: Hall-type response current induced by S_x(z,t)} were evaluated using the conservative choice $\alpha_5=\alpha$, and therefore provide a minimal estimate.

\emph{Summary and Outlook.\textemdash} 
We have shown that excitonic mass does not simply destroy Weyl transport; it creates a new transport regime. In a Weyl excitonic insulator, the mass term reconstructs the low-energy band geometry so that a chiral chemical-potential imbalance $\mu_5\neq0$ generates a Berry structure absent in both ordinary gapped systems and gapless Weyl semimetals. A dynamical axial potential, naturally produced by a transverse sound wave, then converts this reconstructed Berry geometry into a dissipative Hall current. This effect has three distinctive features. First, it is driven by an axial field rather than an electromagnetic vector potential. Second, it exists only in the massive excitonic phase and therefore provides a direct transport signature of interaction-generated Weyl mass. Third, its dependence on the mass, frequency, and $\mu_5$ offers a way to distinguish it experimentally from conventional anomalous transport. Importantly, the response is not encoded in the usual topological $\theta$ term, but emerges from the mass-dependent terms of the effective Breit-type Hamiltonian. Our results therefore reveal sound-induced Hall transport as a probe of Weyl excitonic order and demonstrate that gapped Weyl materials can host axial-field responses beyond the standard anomaly paradigm.

\emph{Acknowledgments\textemdash} 
The authors are essentially grateful to T. H. Hansson for extensive technical discussions and critical reading of the manuscript. This work was supported by National Natural Science Foundation of China (NSFC) under Grant No. 12374332, Quantum Science and Technology-National Science and Technology Major Project No. 2021ZD0301900, Cultivation Project of Shanghai Research Center for Quantum Sciences Grant No.LZPY2024, and Shanghai Science and Technology Innovation Action Plan Grant No. 24LZ1400800. 
\bibliography{ref}
\clearpage
\onecolumngrid
\section{Supplementary Material}
In this supplemental material, we provide the detailed derivation of the semi-classical transport calculations \cite{Son2013,Niu1999,Medel2023} presented in the main text. The structure of this supplemental material is as follows. We review the (3+1)d effective Hamiltonian in Sec. 1 and clarify the non-perturbed Hamiltonian, its eigenvalues, its eigenvectors and the corresponding Berry curvature in Sec. 2. We calculate the wavepacket energy up to $O(\epsilon\delta)$ order in Sec. 3 and solve the modified Boltzmann equation derived from the wavepacket's EOMs in Sec. 4. Lastly, we reach the final expressions of density fluctuation and response current in linear response structure in Sec. 5.

% ===================================================================
\section{1: Effective Hamiltonian for the (3+1)d massive Dirac system}\label{Effective Hamiltonian for the (3+1)d massive Dirac system}
% ===================================================================
We focus on the (3+1)d massive Dirac system with both vector and axial potentials $\{A_\mu,{A^5_\mu}\}$, where the effective Hamiltonian up to $O(m^{-2})$ is obtained by integrating out the negative energy components and rescaling the fields as Ref. \onlinecite{longpaper}:
\be{H_eff_supp_0}
\hat H^{(0)}_{eff} &=& eA_0+\eta {A^5_i}\sigma^i\nonumber\\
\hat H^{(1)}_{eff} &=& \frac{1}{2m}(\hat{\bm{\pi}}^2+\eta^2 ({A^5_0})^2)+\frac{1}{2m}[-eB_i+\eta(\hat\pi_i {A^5_0}+{A^5_0}\hat\pi_i)]\sigma^i\nonumber\\
\hat H^{(2)}_{eff} &=& \frac{e}{8m^2}[-\bm{\nabla}\cdot\mathbf{E}+(\bm{\hat\pi}\times\mathbf{E}-\mathbf{E}\times\bm{\hat\pi})\cdot\bm{\sigma}]-\left(\frac{\eta}{4m^2}\right)(\mathbf{B_5}\cdot\bm{\hat\pi})\nonumber\\
&-&\left(\frac{\eta}{8m^2}\right)[(\hat\pi_i^2{A^5_j}+2\hat\pi_i{A^5_j}\hat\pi_i+{A^5_j}\hat\pi_i^2)-2(\hat\pi_i{A^5_i}\hat\pi_j+\hat\pi_j {A^5_i}\hat\pi_i)]\sigma^j\nonumber\\
&+&\frac{\eta^2}{4m^2}[\epsilon_{ijk}(\partial_i{A^5_0}){A^5_j}\sigma^k]+\frac{e\eta}{2m^2}(B_i{A^5_i})
\ee
where $\hat\pi_i=\hat p_i+eA_i$, $B_i=-\epsilon_{ikl}\partial_k A_l$, $E_i=\partial_0 A_i-\partial_i A_0$, $B_{5,i}=-\epsilon_{ijk}\partial_j {A^5_k}$ \cite{Bertlmann2000}, and $\mu_5\equiv\eta{A^5_0}$ is the constant chiral chemical potential.

In the main text, we consider the regime where all applied potentials $\{{A^5_i},A_\mu\}$ are small $(O(\varepsilon))$ and vary slowly in space-time $(O(\delta))$. Under these assumptions, the effective Hamiltonian is truncated to $O(\varepsilon\delta)$ order:
\be{H_eff_supp_1}
\hat H^{(0)}_{eff} &=& eA_0+\eta {A^5_i}\sigma^i\nonumber\\
\hat H^{(1)}_{eff} &\simeq& \frac{1}{2m}\hat{\bm{\pi}}^2+\frac{1}{2m}(-eB_i+2\mu_5\hat\pi_i)\sigma^i\nonumber\\
\hat H^{(2)}_{eff} &\simeq& \frac{e}{8m^2}[-\bm{\nabla}\cdot\mathbf{E}+(\bm{\hat\pi}\times\mathbf{E}-\mathbf{E}\times\bm{\hat\pi})\cdot\bm{\sigma}]-\left(\frac{\eta}{4m^2}\right)(\mathbf{B_5}\cdot\bm{\hat\pi})\nonumber\\
&-&\left(\frac{\eta}{8m^2}\right)[(\hat\pi_i^2{A^5_j}+2\hat\pi_i{A^5_j}\hat\pi_i+{A^5_j}\hat\pi_i^2)-2(\hat\pi_i{A^5_i}\hat\pi_j+\hat\pi_j {A^5_i}\hat\pi_i)]\sigma^j
\ee

In the following calculation, we will show that the second line of $\hat H_{eff}^{(2)}$ doesn't contribute to the first order wavepacket calculation and this is the reason why we omit it in the main text.

% ===================================================================
\section{2: Unperturbed Hamiltonian, eigenstates, and Berry curvature}\label{Unperturbed Hamiltonian, eigenstates, and Berry curvature}
% ===================================================================

It is useful to separate $\hat H_{\mathrm{eff}}$ into an unperturbed part $\hat H_0$ and perturbation terms. We choose
\be{H0_supp}
\hat H_0(\bm{\pi})=\frac{\bm{\pi}^2}{2m}+\frac{\mu_5}{m}(\bm{\pi}\cdot\bm{\sigma})
\ee
as the unperturbed Hamiltonian. This choice is justified for realistic parameter regimes where the term ${\eta\hbar v_FA^5_i}/{(\mu_5p_s^*/m)}\lesssim O(10^{-4})$, as discussed in the main text. The remaining terms in $\hat H_{\mathrm{eff}}$ are treated as perturbations $\hat H_1 = \hat H_{\mathrm{eff}} - \hat H_0$.

$\hat H_0(\mathbf{p})$ has the following eigenstates and eigenvalues:
\be{eigen_supp}
\hat H_0(\mathbf{p})|\psi_{s}(\mathbf{p})\rangle=\left(\frac{|\mathbf{p}|^2}{2m}+s\cdot\frac{\mu_5|\mathbf{p}|}{m}\right)|\psi_{s}(\mathbf{p})\rangle=\epsilon^{(0)}_s(\mathbf{p})|\psi_{s}(\mathbf{p})\rangle,\quad s\in\{+,-\}
\ee
The two bands are labeled by $s=\pm1$. The corresponding Berry connection is
${\cal A}_{s}(\mathbf{p})=-i\langle\psi_{s}(\mathbf{p})|\bm{\nabla}_{\mathbf{p}}|\psi_s(\mathbf{p})\rangle$,
from which the Berry curvature follows:
\be{Berry_supp}
\bm{\Omega}_s(\mathbf{p})=\bm{\nabla}_{\mathbf{p}}\times{\cal A}_s(\mathbf{p})=s\cdot\frac{\widehat{\mathbf{p}}}{2|\mathbf{p}|^2},\quad s\in\{+,-\}
\ee
The unperturbed band velocity is
\be{vcs_supp}
\mathbf{v}_{cs} \equiv \frac{\partial\epsilon_{s}^{(0)}}{\partial\mathbf{p}_c}
= \frac{(p_c+s\mu_5)\widehat{\mathbf{p}}_c}{m}
\ee

% ===================================================================
\section{3: Wavepacket energy}\label{Wavepacket energy}
% ===================================================================

In this section we present the detailed derivation of the wavepacket energy $\epsilon_s(\mathbf{r}_c,{\mathbf{p}_c})$, which consists of three contributions \cite{Niu1999}:
\be{three_parts_supp}
\epsilon_s(\mathbf{r}_c,{\mathbf{p}_c})=\epsilon^{(0)}_s({\mathbf{p}_c})+\Delta\epsilon^{(0)}_s(\mathbf{r}_c,{\mathbf{p}_c})+\epsilon^{(1)}_s(\mathbf{r}_c,{\mathbf{p}_c})
\ee
where $\epsilon^{(0)}_s({\mathbf{p}_c})={|\mathbf{p}_c|}^2/2m + s\mu_5 {|\mathbf{p}_c|}/m$ is the unperturbed band energy. To evaluate the corrections, we recall the definition of a wavepacket $|w_s\rangle$:
\be{wp_def_supp}
|w_s\rangle=\int d^3p\;a(\mathbf{p},t)|\psi_s(\mathbf{p})\rangle=\int d^3p\;a(\mathbf{p},t)e^{i\mathbf{p}\cdot\mathbf{\hat r}}|u_{s}(\mathbf{p})\rangle
\ee

The envelope function $a(\mathbf{p},t)$ is sharply peaked around the center momentum ${\mathbf{p}_c}$, and the wavepacket is centered at position $\mathbf{r}_c$.

\subsection{Calculation of $\Delta\epsilon_s^{(0)}(\mathbf{r}_c,{\mathbf{p}_c})$}

The correction $\Delta\epsilon_s^{(0)}(\mathbf{r}_c,\mathbf{p}_c)$ arises from the spatial gradient of the unperturbed Hamiltonian evaluated at the wavepacket center. In the wavepacket formalism, it is given by:
\be{calculation of Delta-epsilon_s^(0)}
\Delta\epsilon_s^{(0)}(\mathbf{r}_c,{\mathbf{p}_c}) &=& \langle w_s|\frac{1}{2}\left[(\mathbf{\hat r}-\mathbf{r}_c)\cdot\frac{\partial \hat H_0}{\partial\mathbf{r}}|_{\mathbf{r}=\mathbf{r_c}}+\frac{\partial\hat H_0}{\partial\mathbf{r}}|_{\mathbf{r}=\mathbf{r_c}}\cdot(\mathbf{\hat r}-\mathbf{r_c})\right]|w_s\rangle\nonumber\\
&=&\frac{i}{2}\left\{\langle\frac{\partial u_s}{\partial p_{ci}}|\hat H_0-\epsilon_s({{|\mathbf{p}_c|}})|\frac{\partial u_s}{\partial p_{cj}}\rangle(e\partial_i A_j)+\langle\frac{\partial u_s}{\partial p_{ci}}|\hat H_0-\epsilon_{s}({|\mathbf{p}_c|})|\frac{\partial u_s}{\partial p_{cj}}\rangle(-e\partial_j A_i)\right\}\nonumber\\
&=&\frac{-eB_k}{2}\left\{i\epsilon^{ijk}\langle\frac{\partial u_s}{\partial p_{ci}}|\hat H_0-\epsilon_{s}({|\mathbf{p}_c|})|\frac{\partial u_s}{\partial p_{cj}}\rangle\right\}\nonumber\\
&=&\frac{-e\mu_5 B_k}{2m}\left\{i\epsilon^{ijk}\langle\frac{\partial u_s}{\partial p_{ci}}|{\mathbf{p}_c}\cdot\bm{\sigma}-s\cdot {|\mathbf{p}_c|}|\frac{\partial u_s}{\partial p_{cj}}\rangle\right\}\nonumber\\
&=&\frac{-e\mu_5 (\mathbf{B}\cdot{\mathbf{p}_c})}{2m{|\mathbf{p}_c|}^2}
\ee

$\Delta\epsilon_s^{(0)}(\mathbf{r}_c,\mathbf{p}_c)$ will contribute to the dynamical modification of the chiral magnetic effect (CME) in Eq.\eqref{final expression of the epsilon-delta order response current} later.

\subsection{Calculation of $\epsilon_s^{(1)}(\mathbf{r}_c,{\mathbf{p}_c})$}

The first-order perturbation energy is:
\be{eps1_supp}
\epsilon_s^{(1)}(\mathbf{r}_c,{\mathbf{p}_c})=\langle w_s|\hat H_1(\mathbf{r}_c,{\mathbf{p}_c})|w_s\rangle=\langle u_{s}|\hat H_1(\mathbf{r}_c,2{\mathbf{p}_c})|u_s\rangle
\ee

Collecting all perturbation terms from Eq.\eqref{H_eff_supp_1}, the explicit expression for $\hat H_1(\mathbf{r}_c,2{\mathbf{p}_c})$ reads:
\be{H1_supp}
\hat H_1(\mathbf{r}_c,2{\mathbf{p}_c})&=&\left[\eta {A^5_j}-\frac{e}{2m}B_j+\frac{e}{2m^2}\epsilon^{ikj}p_{ci}E_k+\frac{2\eta}{m^2}{|\mathbf{p}_c|}^2{A^5_j}-\frac{2\eta}{m^2}(p_{ci}{A^5_i})p_{cj}\right]\sigma^j\nonumber\\
&+&\left(-\frac{\eta}{2m^2}B_{5i}p_{ci}\right)\nonumber\\
&\equiv&h_j(\mathbf{r}_c,{\mathbf{p}_c})\sigma^j+h_0(\mathbf{r}_c,{\mathbf{p}_c})
\ee

Note that in this step we localize both ${A^5_i}$ and $A_\mu$ at $\mathbf{r}_c$, so they commute with momentum operators. Using the spin expectation value $\langle u_{s}({\mathbf{p}_c})|\bm{\sigma}|u_{s}({\mathbf{p}_c})\rangle=s\cdot\widehat{\mathbf{p}}_c$, we obtain:
\be{eps1_final_supp}
\epsilon_{s}^{(1)}(\mathbf{r}_c,{\mathbf{p}_c})=s\left(\frac{\eta {A^5_j}p_{cj}}{{|\mathbf{p}_c|}}-\frac{eB_jp_{cj}}{2m{|\mathbf{p}_c|}}\right)-\left(\frac{\eta}{2m^2}B_{5j}p_{cj}\right)
\ee

Here we can see that the second line of $\hat H_{eff}^{(2)}$ doesn't contribute to the first order wavepacket calculation as claimed before. Combining all three contributions, the wavepacket energy for $\hat H_{\mathrm{eff}}$ is:
\be{eps_total_supp}
\epsilon_{s}(\mathbf{r}_c,{\mathbf{p}_c})=\left(\frac{{|\mathbf{p}_c|}^2}{2m}+\frac{s\mu_5{|\mathbf{p}_c|}}{m}\right)+\left(\frac{-e\mu_5B_jp_{cj}}{2m{|\mathbf{p}_c|}^2}\right)+s\left(\frac{\eta {A^5_j}p_{cj}}{{|\mathbf{p}_c|}}-\frac{eB_jp_{cj}}{2m{|\mathbf{p}_c|}}\right)-\left(\frac{\eta}{2m^2}B_{5j}p_{cj}\right)\nonumber\\
\ee

% ===================================================================
\section{4: Boltzmann transport equation and distribution function}\label{Boltzmann transport equation and distribution function}
% ===================================================================

\subsection{Semi-classical equations of motion and the modified Boltzmann equation}

The dynamics of a wavepacket in phase space are governed by the semi-classical equations of motion \cite{Niu1999,Son2013}:
\be{EOM_supp}
\mathbf{\dot p_c}&=&-\frac{\partial\epsilon_s}{\partial\mathbf{r_c}}+e(\mathbf{E}+\mathbf{\dot r_c}\times\mathbf{B})\label{EOM1_supp}\\
\mathbf{\dot r_c}&=&\frac{\partial\epsilon_s}{\partial\mathbf{p_c}}+\mathbf{\dot p_c}\times\bm{\Omega}_s\label{EOM2_supp}
\ee

These equations incorporate both the Lorentz force from electromagnetic fields, mass dependent modification of electromagnetic fields from axial potentials and the anomalous velocity from Berry curvature. The wavepacket energy $\epsilon_s(\mathbf{r}_c,\mathbf{p}_c)$ is given by Eq.\eqref{eps_total_supp}. The electron distribution function $n_s(\mathbf{r}_c,\mathbf{p}_c,t)$ satisfies the Boltzmann equation:
\be{Boltz_supp}
\partial_t n_s + {\bf \dot{r}}_c\cdot\partial_{\bf {r}_c}n_s +\dot{\mathbf{p}}_c\cdot\partial_{\mathbf{p}_c}n_s = I[n_s]
\ee
where $I[n_s]$ is the collision integral. Substituting the EOMs Eqs.(\ref{EOM1_supp})-(\ref{EOM2_supp}) into the Boltzmann equation and solving for $\mathbf{\dot r}_c$ and $\mathbf{\dot p}_c$, we obtain the modified Boltzmann equation:
\be{Boltz_mod_supp}
\partial_t n_s &+& \frac{1}{1+e\mathbf{B}\cdot\bm{\Omega}_s}\left[\frac{\partial\epsilon_s}{\partial\mathbf{p}_c}+e\tilde{\mathbf{E}}_{s}\times\bm{\Omega}_s+e\left(\frac{\partial\epsilon_s}{\partial\mathbf{p}_c}\cdot\bm{\Omega}_s\right)\mathbf{B}\right]\cdot\frac{\partial n_s}{\partial\mathbf{r_c}}\nonumber\\
&+&\frac{1}{1+e\mathbf{B}\cdot\bm{\Omega}_s}\left[e\tilde{\mathbf{E}}_{s}+e\frac{\partial\epsilon_s}{\partial\mathbf{p}_c}\times\mathbf{B}+e^2(\tilde{\mathbf{E}}_{s}\cdot\mathbf{B})\bm{\Omega}_s\right]\cdot\frac{\partial n_s}{\partial\mathbf{p}_c}=I[n_s]
\ee

We define the modified electric field $\tilde{\mathbf{E}}_{s}=\mathbf{E}-e^{-1}\cdot{\partial\epsilon_s}/{\partial\mathbf{r}_c}$, which accounts for the spatial variation of the wavepacket energy due to inhomogeneous external potentials.

\subsection{Order expansion of the distribution function and the Boltzmann equation}

Since the applied potentials are small $(O(\varepsilon))$ and vary slowly $(O(\delta))$, we expand the distribution function to $O(\varepsilon\delta)$ order:
\be{n_expand_supp}
n_s(\mathbf{r}_c,\mathbf{p}_c,t)&\simeq&\Theta(\mu_s-\epsilon_s)+\delta n_s^{(\varepsilon)}+\delta n_s^{(\varepsilon\delta)}\nonumber\\
&=&\Theta(\mu_s-\epsilon_s^{(0)})+\left[\delta n_s^{(\varepsilon)}-s\eta {A^5_j}\widehat{p}_{cj}\,\delta(\mu_s-\epsilon_s^{(0)})\right]\nonumber\\
&+&\left[\delta n_s^{(\varepsilon\delta)}+\left(\frac{seB_j\widehat{p}_{cj}}{2m}+\frac{\eta B_{5j}p_{cj}}{2m^2}+\frac{e\mu_5B_j\widehat{p}_{cj}}{2mp_c}\right)\delta(\mu_s-\epsilon_s^{(0)})\right]
\ee

Here $\mu_s=\mu+s\mu_5$ is the effective chemical potential for the $s$-band, and $\Theta(x)$ is the Heaviside step function. The equilibrium distribution is the zero temperature Fermi-Dirac distribution $\Theta(\mu_s-\epsilon_s)$, which we expand around the unperturbed energy $\epsilon_s^{(0)}=\epsilon_s(\mathbf{r}_c,\mathbf{p}_c)|_{\{A_\mu,{A^5_i}\}\to 0}$.

Keeping only terms of the relevant order in the modified Boltzmann equation, and describing the relaxation process with a single relaxation time $\tau_0$ via $I[n_s]=-\tau_0^{-1}[\delta n_s^{(\varepsilon)}+\delta n_s^{(\varepsilon\delta)}]$, we obtain:
\be{Boltz_simple_supp}
(\partial_t+\mathbf{v}_{cs}\cdot\partial_{\mathbf{r}_c})n_s+\left(e\mathbf{E}-\frac{\partial\epsilon_s}{\partial\mathbf{r}_c}+e\mathbf{v}_{cs}\times\mathbf{B}\right)\cdot\frac{\partial n_s}{\partial\mathbf{p}_c}=-\frac{1}{\tau_0}[\delta n_s^{(\varepsilon)}+\delta n_s^{(\varepsilon\delta)}]
\ee

Expanding Eq.\eqref{Boltz_simple_supp} order by order in $\varepsilon$ and $\delta$ yields the kinetic equations for $\delta n_s^{(\varepsilon)}$ and $\delta n_s^{(\varepsilon\delta)}$:
\be{kinetic_order_supp}
(\partial_t+\mathbf{v}_{cs}\cdot\partial_{\mathbf{r}_c}+\tau_0^{-1})\delta n_s^{(\varepsilon)}&=&\delta(\mu_s-\epsilon_s^{(0)})[s\eta\dot{A}^5_j\widehat{p}_{cj}+e{\mathbf{E}}\cdot\mathbf{v}_{cs}]\label{kin_eps}\\
(\partial_t+\mathbf{v}_{cs}\cdot\partial_{\mathbf{r}_c}+\tau_0^{-1})\delta n_s^{(\varepsilon\delta)}&=&\delta(\mu_s-\epsilon_s^{(0)})\Big[-\frac{e\mu_5\dot{B}_j\widehat{p}_{cj}}{2mp_c}-\frac{\eta\dot{B}_{5j}p_{cj}}{2m^2}-\frac{se\dot{B}_j\widehat{p}_{cj}}{2m}\Big]\label{kin_epsdelta}
\ee

The right-hand sides of these equations arise from the spatial and temporal derivatives of the external potentials acting on the equilibrium distribution.

\subsection{Solution via the characteristic line method}

Eqs.(\ref{kin_eps})-(\ref{kin_epsdelta}) are first-order linear partial differential equations that can be solved by the method of characteristics. The characteristic lines are $\mathbf{r}_c(t)=\mathbf{r}_c(0)+\mathbf{v}_{cs}t$, along which the solution takes the form of a retarded-time integral:
\be{dn_solution_supp}
\delta n_s^{(\varepsilon)}&=&\int_0^{+\infty}d\tau\,e^{-\tau_0^{-1}\tau}\,\delta(\mu_s-\epsilon_s^{(0)})\,\big[s\eta\dot{A}^5_j\widehat{p}_{cj}+e{\mathbf{E}}\cdot\mathbf{v}_{cs}\big]\label{dn_eps_supp}\\
\delta n_s^{(\varepsilon\delta)}&=&\int_0^{+\infty}d\tau\,e^{-\tau_0^{-1}\tau}\,\delta(\mu_s-\epsilon_s^{(0)})\,\Big[-\frac{e\mu_5\dot{B}_j\widehat{p}_{cj}}{2mp_c}-\frac{\eta\dot{B}_{5j}p_{cj}}{2m^2}-\frac{se\dot{B}_j\widehat{p}_{cj}}{2m}\Big]\label{dn_epsdelta_supp}
\ee

In these expressions, the potentials are evaluated along the characteristic:
\be{shorthand_supp}
{A^5_j}&\equiv&{A^5_j}(\mathbf{r}_c-\tau\mathbf{v}_{cs},t-\tau)={A^5_j}(\tilde r_c-\tau\tilde v_{cs})\\
A_\mu&\equiv&A_\mu(\mathbf{r}_c-\tau\mathbf{v}_{cs},t-\tau)=A_\mu(\tilde r_c-\tau\tilde v_{cs})
\ee

This retarded-time structure encodes the causal response of the electron distribution to time-dependent external perturbations, with the exponential factor $e^{-\tau_0^{-1}\tau}$ describing relaxation due to scattering.

% ===================================================================
\section{5: Response density fluctuation and current densities}\label{Response density fluctuation and current densities}
% ===================================================================

In the wavepacket formalism, the total current carried by band $s$ is \cite{Son2013,Gorbar2017}:
\be{j_general_supp}
\mathbf{j}_s(\mathbf{r}_c,t)=e\int\frac{d^3p_c}{(2\pi)^3}\left[-\epsilon_s\frac{\partial n_s}{\partial\mathbf{p}_c}-e\left(\bm{\Omega}_s\cdot\frac{\partial n_s}{\partial\mathbf{p}_c}\right)\epsilon_s\mathbf{B}-\epsilon_s\left(\bm{\Omega}_s\times\frac{\partial n_s}{\partial\mathbf{r}_c}\right)\right]
\ee

The three terms correspond respectively to: (i) the conduction current from the band velocity, (ii) the chiral magnetic effect (CME) contribution, and (iii) the anomalous Hall effect from the Berry curvature in real space. 

Expanding the current by order in $\varepsilon$ and $\delta$, and performing a Fourier transform to momentum-frequency space, we obtain the response density fluctuation and current densities:
\be{j_order_supp}
\delta\rho_s^{(\varepsilon)}(\mathbf{q},\omega)&=&e\int dt\,d^3r_c\,e^{i\omega t-i\mathbf{q}\cdot\mathbf{r}_c}\int\frac{d^3p_c}{(2\pi)^3}\,\delta n_s^{(\varepsilon)}(\mathbf{r}_c,\mathbf{p}_c,t)\\
\mathbf{j}_s^{(\varepsilon)}(\mathbf{q},\omega)&=&\int dt\,d^3r_c\;e^{i\omega t-i\mathbf{q}\cdot\mathbf{r}_c}\int\frac{d^3p_c}{(2\pi)^3}\,\delta n_s^{(\varepsilon)}(\mathbf{r}_c,\mathbf{p}_c,t)\cdot\mathbf{v}_{cs}\\
\mathbf{j}_s^{(\varepsilon\delta)}(\mathbf{q},\omega)&=&\int dt\,d^3r_c\;e^{i\omega t-i\mathbf{q}\cdot\mathbf{r}_c}\int\frac{d^3p_c}{(2\pi)^3}\,
\bigg[\delta n_s^{(\varepsilon\delta)}\cdot\mathbf{v}_{cs}
-e\Big(\bm{\Omega_s}\cdot\frac{\partial n^{(0)}_s}{\partial\mathbf{p_c}}\Big)\epsilon_s^{(0)}\mathbf{B}-\epsilon_s^{(0)}\Big(\bm{\Omega}_s\times\frac{\partial\delta n_s^{(\varepsilon)}}{\partial\mathbf{r}_c}\Big)\bigg]\nonumber\\
\ee

\subsection{Fourier transform and Fermi surface integrals}

To evaluate these expressions, we insert the distribution function solutions Eqs.\eqref{dn_eps_supp}-\eqref{dn_epsdelta_supp} into the current formulas. The key technical step is the Fourier transform of the retarded-time integral:
\be{FT_supp}
\mathcal{F}[f(\tilde r_c-\tau\tilde v_{cs})]&\equiv&\int dt\,d^3r_c\;e^{i\omega t-i\mathbf{q}\cdot\mathbf{r}_c}\int_0^{+\infty}d\tau\;e^{-\tau_0^{-1}\tau}f(\tilde r_c-\tau\tilde v_{cs})\nonumber\\
&=&\frac{i\,f(\mathbf{q},\omega)}{\omega-\mathbf{q}\cdot\mathbf{v}_{cs}+i\tau_0^{-1}}
\ee

The momentum-space integration over the $\delta$-function constraint $\delta(\mu_s-\epsilon_s^{(0)})$ restricts the integration to the Fermi surface, reducing the 3D momentum integral to an angular integral over the Fermi surface:
\be{I_integrals_supp}
I_{0j,s}(\mathbf{q},\omega)&=&\int\frac{d\widehat{\mathbf{p}}_c}{(2\pi)^3}\frac{\widehat{p}_{cj}}{\omega-v_s^{*}\,\mathbf{q}\cdot\widehat{\mathbf{p}}_c+i\tau_0^{-1}}\\
I_{ij,s}(\mathbf{q},\omega)&=&\int\frac{d\widehat{\mathbf{p}}_c}{(2\pi)^3}\frac{\widehat{p}_{ci}\widehat{p}_{cj}}{\omega-v_s^{*}\,\mathbf{q}\cdot\widehat{\mathbf{p}}_c+i\tau_0^{-1}}
\ee

Here we have introduced the effective wavepacket momentum $p_s^{*}=-s\mu_5+\sqrt{\mu_5^2+2m\mu_s}$ and effective wavepacket velocity $v_s^{*}=\sqrt{\mu_5^2+2m\mu_s}/{m}$, which characterize the Fermi surface of the $s$-band.

\subsection{Final expressions for the response density fluctuation and current densities}

Carrying out the momentum integrals, we obtain the final density fluctuation and response current expressions:
\be{j_final_supp}
\delta\rho_s^{(\varepsilon)}(\mathbf{q},\omega)&=&(ie^2{p_s^*}^2)\,I_{0j,s}(\mathbf{q},\omega)\,E_{s,j}^{(\varepsilon)}\\
j_{s,i}^{(\varepsilon)}(\mathbf{q},\omega)&=&(ie^2{p^*_s}^2v^*_s)\,I_{ij,s}(\mathbf{q},\omega)\,E_{s,j}^{(\varepsilon)}\\
j_{s,i}^{(\varepsilon\delta)}(\mathbf{q},\omega)&=&\left[\frac{s\mu_s e^2B^{(0)}_i}{4\pi^2}-\frac{e^2{p^*_s}\mu_5}{2m}\,\omega\,I_{ij,s}(\mathbf{q},\omega)B^{(0)}_j\right]\nonumber\\
&-&\frac{se^2{p^*_s}^2}{2m}\,\omega\,I_{ij,s}(\mathbf{q},\omega)B^{(\varepsilon\delta)}_{s,j}\nonumber\\
&+&\Theta(|\mu_5|)\,se^2\mu_s\,\epsilon_{ijk}q_k\,I_{jl,s}(\mathbf{q},\omega)E^{(\varepsilon)}_{s,l}\label{final expression of the epsilon-delta order response current}
\ee

where the effective electromagnetic fields experienced by the wavepacket are:
\be{EM_eff_supp}
E_{s,i}^{(\varepsilon)}&=&-i\left[\omega\left(\frac{s}{v_s^{*}}{A^5_i}+A_i\right)+q_iA_0\right]=E_i^{(0)}-\frac{i\omega s}{v_s^*}A^5_i\\
B_{s,i}^{(\varepsilon\delta)}&=&-i\epsilon_{ikl}q_k\left(\frac{sp^*_s}{m}{A^5_l}+A_l\right)=B_i^{(0)}-i\epsilon_{ikl}q_k\cdot\frac{sp_s^*}{m}A_l^5
\ee

The standard electromagnetic fields are $E_i^{(0)}=-i(\omega A_i+q_i A_0)$ and $B_i^{(0)}=-i\epsilon_{ikl}q_k A_l$. The modifications proportional to ${A^5_i}$ represent the mass-dependent coupling between axial and vector potentials in the gapped system, which is distinct from the simple chiral shift $A_\mu^{\pm}\equiv A_\mu\pm{A^5_\mu}$ found in gapless Weyl semimetals.

The three lines in $j_{s,i}^{(\varepsilon\delta)}$ have distinct physical origins: the first line contains the chiral magnetic effect and its dynamical massive correction from $\Delta\epsilon_s^{(0)}$, the second line contains the current driven by dynamical axial potentials via the $(-\eta/4m^2)(\mathbf{B}_5\cdot\hat{\bm{\pi}})$ term in $\hat H_{\mathrm{eff}}^{(2)}$, and the third line represents the anomalous Hall effect arising from the spin-orbit interaction between $(\eta{A^5_i}\sigma^i)$ in $\hat H_{\mathrm{eff}}^{(0)}$ and the Berry curvature $\bm{\Omega}_s$, which is present only when $\mu_5\neq0$ (enforced by $\Theta(|\mu_5|)$).

%\bibliography{sup_ref}
\end{document}